\documentstyle[emulateapj,graphicx]{article}
\def\gsim{\;\rlap{\lower 2.5pt
 \hbox{$\sim$}}\raise 1.5pt\hbox{$>$}\;}
\def\lsim{\;\rlap{\lower 2.5pt
   \hbox{$\sim$}}\raise 1.5pt\hbox{$<$}\;}

\newcommand{\beq}{\begin{equation}}
\newcommand{\eeq}{\end{equation}}
\def\myputfigure#1#2#3#4#5%
{\hskip0.03\textwidth\vskip#5pt
\makebox[0pt]{\hskip#2in
\includegraphics[width=#3\textwidth]{#1}}\vskip#4pt\hfill}


\lefthead{Haiman \& Rees}
\righthead{Extended Ly$\alpha$ Emission Around Young Quasars}

\begin{document}
\title{Extended Ly$\alpha$ Emission Around Young 
Quasars: a Constraint on Galaxy Formation}
\author{Zolt\'an Haiman\altaffilmark{1}} \affil{Princeton University
Observatory, Princeton, NJ 08544, USA\\ zoltan@astro.princeton.edu}
\vspace{0.5\baselineskip}
\author{Martin J. Rees}
\vspace{0.1\baselineskip}
\affil{Institute of Astronomy, Madingley Road, Cambridge, CB3 0HA, UK\\
mjr@ast.cam.ac.uk}
\altaffiltext{1}{Hubble Fellow}

\vspace{\baselineskip}
\begin{abstract}
The early stage in the formation of a galaxy inevitably involves a spatially
extended distribution of infalling, cold gas.  If a central luminous quasar
turned on during this phase, it would result in significant extended Ly$\alpha$
emission, possibly accompanied by other lines.  For halos condensing at
redshifts $3\lsim z \lsim 8$ and having virial temperatures $2\times 10^5~{\rm
K}\lsim T_{\rm vir}\lsim 2\times 10^6~{\rm K}$, this emission results in a
``fuzz'' of characteristic angular diameter of a few arcseconds, and surface
brightness $\sim 10^{-18}-10^{-16}~{\rm erg~s^{-1}~cm^{-2}~asec^{-2}}$.  The
fuzz around bright, high redshift quasars could be detected in deep narrow band
imaging with current telescopes, providing a direct constraint on galaxy
formation models.  The absence of detectable fuzz might suggest that most that
most of the protogalaxy's gas settles to a self--gravitating disk before a
quasar turns on. However, continued gas infall from large radii, or an
on--going merger spreading cold gas over a large solid angle, during the
luminous quasar phase could also result in extended Ly$\alpha$ emission, and
can be constrained by deep narrow band imaging.
\end{abstract}
\keywords{cosmology: theory -- galaxies: formation -- quasars: general --
black hole physics}

\section{Introduction}
\label{sec:introduction}

There has been considerable progress recently in the study of the
formation of galaxies and quasars.  Both the galaxy (Steidel et
al. 1999) and quasar (Fan et al. 2000a) luminosity functions are now
observationally determined to redshifts of $z\sim 4$, probing into the
epoch when at least some of these objects are still expected to be
assembling.  Although the evolutions of the galaxy and quasar
populations is generally expected to be connected, the detailed nature
of this link is yet to be elucidated.  One possibility is that quasars
represent a brief phase in the early life of each galaxy.  Indeed,
theoretically, one might expect quasar activity to be triggered by the
mergers that a galaxy is experiencing during its assembly
(e.g. Cavaliere \& Vittorini 2000, Kauffmann \& Haehnelt 2000).  This
assertion has some observational support: the galaxy abundance appears
to peak at a somewhat later redshift than does the quasar abundance
(e.g. Madau \& Pozzetti 2000); and at least infrared--selected quasars
appear to preferentially reside in morphologically disturbed hosts
(e.g. Baker \& Clements 1997).

It is therefore interesting to consider a quasar that turns on within an
assembling protogalaxy.  The behavior of the gas inside a dark matter (DM)
condensation during the formation of disk galaxies has been investigated in
semi--analytic schemes (White \& Rees 1978; Fall \& Efstathiou 1980; Mo, Mao \&
White 1998; van den Bosch \& Dalcanton 2000), and numerical simulations (Katz
1991; Navarro \& White 1994; Navarro, Frenk \& White 1997 [NFW]; Navarro \&
Steinmetz 1997; Moore et al. 1999). In these studies, the bulk of the baryons
in the halo cool and settle into a rotationally supported disk.  In the
simplest picture, the disk material originates as smooth gas, collapsing from
the virial radius $R_{\rm vir}$ to its final orbital radius of $\sim \lambda
R_{\rm vir}$, where $\lambda\sim 0.05$ is the typical spin parameter. Numerical
simulations have revealed a more complex process, where a fraction of the
infalling gas forms smaller clumps early on; these clumps then progressively
merge together, collide, and dissipate to form larger systems.

A robust feature of galaxy formation is, at least in the early stages, a
spatially extended distribution of gas.  At the densities expected at $z>2$ for
this gas ($\gsim 100$ times the background density), the radiative cooling
time--scales are shorter than the typical dynamical times.  In the absence of
heat input from stars or a quasar, a significant fraction of this gas would
therefore be cold ($T\lsim 10^4$K) and neutral (Fall \& Rees 1985).  As it
contracts inside the DM halo, the cold gas is heated by the halo potential;
this heat is dissipated largely via collisional excitation of the Ly$\alpha$,
and possible other (metal) lines.  The resulting line radiation results in a
potentially observable, extended, low--surface brightness Ly$\alpha$ ``fuzz''
(Haiman et al. 2000; Fardal et al. 2000).

The presence of a bright central quasar during this phase could strongly
enhance the surface brightness in the Ly$\alpha$ line, since a significant
fraction of the quasar's ionizing radiation could be reprocessed into
recombination radiation in the same line.  The purpose of this paper is
to quantify the expected Ly$\alpha$ fluxes, and describe constraints that the
presence or absence of extended Ly$\alpha$ ``fuzz'' around luminous quasars
implies for the host galaxy.  This paper is organized as follows: in
\S~\ref{sec:model}, we describe a simple toy--model for the distribution of
cold gas in a protogalaxy; in \S~\ref{sec:lya}, we characterize the
resulting Ly$\alpha$ ``fuzz''; in \S~\ref{sec:constraints}, we discuss
the implied constraints on galaxy formation; and in
\S~\ref{sec:conclusions}, we summarize our conclusions. Throughout
this work, we adopt a $\Lambda$CDM cosmology with $\Omega_m=0.3,
\Omega_b=0.04, \Omega_\Lambda=0.7, h=0.7$, and $\sigma_8=0.9$.

\section{The Amount of Cold Gas: A Spherical Model}
\label{sec:model}

In this section, we describe a simple model for the structure of a
spherically symmetric, two--phase gas in a DM halo without a central
ionizing source.  We shall assume that the gas has a centrally
condensed radial profile $\rho(r)$, and that it consists of a
two--phase medium with a cold ($T\sim 10^4$K) mass fraction $f(r)$,
and a hot ($T\sim T_{\rm vir}$) mass fraction $1-f(r)$.  The physical
state of the gas in a realistic protogalaxy is, of course, likely to
be much more complicated, with asymmetric infall of pre--existing
dense clumps, an the settling of gas in the central regions into a
disk.  Nevertheless, this model will serve as a reference point for
our discussion in \S~\ref{sec:lya} and \S~\ref{sec:constraints} below.

For sake of concreteness, we assume that the gas initially settles to
a radial profile $\rho_{\rm gas}(r)$ that satisfies hydrostatic
equilibrium within a dark matter halo.  For the profile of the halo,
we follow the description in Navarro, Frenk \& White (1997) with a
concentration parameter $c=5$. However, we assume that the average
enclosed mass density is a fraction $\Delta_c$ of the critical
density, where we obtain $\Delta_c\lsim 18\pi^2$ from the spherical
top--hat collapse (rather than using the fixed value of 200).  To
obtain $\rho_{\rm gas}(r)$, we also assume that the gas is isothermal
at the virial temperature of the halo (Makino, Sasaki \& Suto 1998).
It is useful to note that under these assumptions, the gas is
centrally condensed with a flat core, and has a mean internal
(volume--averaged) ``clumping'' of $\langle \rho_{\rm gas}^2 \rangle
\approx 2.8\times 10^6 \rho_{\rm b}^2$ relative to the background
baryon density $\rho_{\rm b}$.  

In order to obtain the density $\rho_{\rm cold}(r)$ of neutral gas, we
next assume that a mass fraction $f(r)$ of the gas cools and condenses
out into a cold phase with $T=10^4$K. Pressure balance with the
remaining hot, ionized gas implies that the densities of the two
components are
\begin{eqnarray}
\nonumber
\eta_{\rm cold}&=&\frac{\rho_{\rm cold}}{\rho_{\rm gas}}
 =f+(1-f)\frac{T_{\rm vir}}{10^4~{\rm K}}\geq 1\\
\eta_{\rm hot}&=&\frac{\rho_{\rm hot}}{\rho_{\rm gas}}
 =(1-f)+f\frac{10^4~{\rm K}}{T_{\rm vir}}\leq 1.
\label{eq:frac}
\end{eqnarray}
The value of $f=f(r)$ is determined from the condition that the
cooling time of the rarefied hot component is equal to the age of the
system.  The cooling time is given by $t_{\rm cool}= (3/2) \mu m_p k_B
T_{\rm vir}/(\rho_{\rm hot}\Lambda)$, where $\Lambda\sim10^{-23}~{\rm
erg~s^{-1}~cm^3}$ is the cooling function at $T_{\rm vir}$.  We
conservatively adopt a metal--free cooling function (B\"ohringer \&
Hensler 1989).  For the age of the system, we adopt $20\%$ of the
Hubble time, $t_{\rm age} =0.2(6\pi G \rho_z)^{-1/2}$, where $\rho_z$
is the total (dark matter + baryons) mean background density at
redshift $z$. This is roughly the ``mass--doubling'' time for halos of
interest in the extended Press--Schechter formalism (see Lacey \& Cole
1993; and Haiman, Spaans \& Quataert 2000 for a discussion).

Illustrative results for the cold fraction $f(r)$ as a function of
radius under the above assumptions are shown in
Figure~\ref{fig:fcold}.  The dashed and solid curves describe halos at
redshifts $z=3$, and $z =5$, respectively.  At both redshifts, four
different halo sizes are shown, with virial temperatures of $T_{\rm
vir}$ = $2\times10^5~{\rm K}$, $4\times10^5~{\rm K}$, $10^6~{\rm K}$,
and $10^7~{\rm K}$ (top to bottom).  At $z=3$, these correspond to
halo masses $M_{\rm halo} \approx$ $4\times10^{10}~{\rm M_\odot}$,
$10^{11}~{\rm M_\odot}$, $5\times10^{11}~{\rm M_\odot}$, and
$10^{13}~{\rm M_\odot}$; at $z=5$, the halo masses are a factor of
$\sim$ two smaller.

As expected, the cold fraction is a function of radius in each case,
increasing towards $r=0$ where densities are higher and cooling times
are shorter. The cold fraction is larger for smaller halos, which have
smaller initial binding energies, and hence cool more rapidly.  The
cold fraction increases with redshift, because of the higher densities
and shorter cooling times.  Although our model is highly idealized, it
captures the above scalings (which are expected to be robust, as long
as the cold fraction is determined by the cooling time), and provides
a conservative estimate of the amount of cold gas.  For comparison, we
note that Mo \& Miralda-Escud\'e (1996) have used a different,
simplified model of a two--phase medium in order to model quasar line
absorption systems.  In their model, nearly all of the gas is cold
within the ``cooling radius'', defined as the radius at which the
cooling time equals the Hubble time (exceeding $R_{\rm vir}$ for all
halos considered here).

\myputfigure{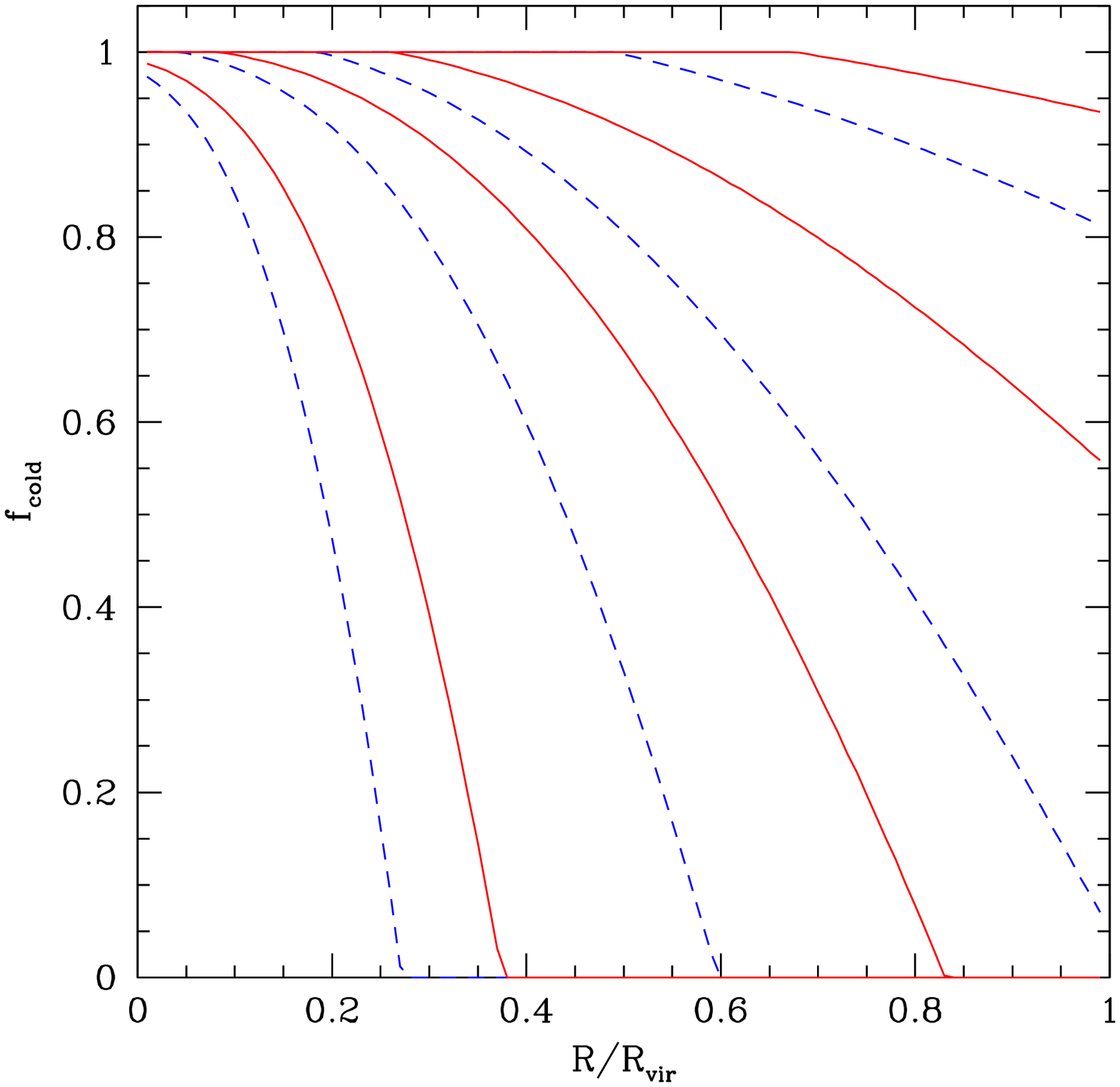}{3.2}{0.45}{-10}{-10} \figcaption{\label{fig:fcold}
The mass fraction $f$ of cold gas as a function of radius for
two--phase gas inside a DM halo with an NFW profile. The dashed curves
correspond to halos at redshift $z=3$, and the solid curves to $z=5$.
At both redshifts, four different halo sizes are shown, corresponding
(top to bottom) to, $T_{\rm vir}$ = $2\times10^5~{\rm K}$,
$4\times10^5~{\rm K}$, $10^6~{\rm K}$, and $10^7~{\rm K}$.}
\vspace{\baselineskip}

Finally, we emphasize that our model assumptions apply only within the
virial radius, where the gas has been shock--heated.  In the rest of
this paper, we focus on the reprocessing of quasar light within
protogalactic halos.  However, the clumpy background gas outside the
virial region can itself reprocess ionizing UV radiation into
Ly$\alpha$ emission.  Indeed, Gould \& Weinberg (1996) have shown that
Ly$\alpha$ absorption systems, illuminated by the UV background, can
cause significant Ly$\alpha$ fluorescence (see Bunker et al. 1998 for
a current status of observations).  Numerical simulations suggest that
the background gas near a collapsing protogalaxy is highly clumped,
and hence could reprocess any ionizing radiation from the protogalaxy
into Ly$\alpha$ emission.  This type of reprocessing of quasar light
could be the interpretation of extended Ly$\alpha$--emitting blobs
found in the vicinity of quasars (Hu et al. 1991; Hu et al. 1996).
Although these blobs are likely associated with small satellite
protogalaxies, their Ly$\alpha$ emission could be dominated by
reprocessed quasar light.

\section{Ionization by a Central Source and the Accompanying Ly$\alpha$ Fuzz}
\label{sec:lya}

We next consider a quasar that turns on at the center of the halo
described above.  In general, the cold gas will then be photoionized
by the quasar's UV radiation, out to a radius that depends on the
quasar's ionizing luminosity.  Utilizing the spatial distribution of
cold gas obtained in our models, we first argue that reasonably--sized
quasar black holes (BHs) can keep most of the cold phase photoionized.
We then compute the characteristic Ly$\alpha$ surface brightness of
such quasar--illuminated protogalaxies.

\subsection{Effect of an Ionizing Source}

The radius of the photoionized Str\"omgren sphere around a central
quasar embedded in a spherical halo is given by
\begin{equation}
R_{\rm HII} = 
\left[ \frac{3\dot N_{\rm ph}}{4\pi\alpha_B\langle n_{\rm H}^2 \rangle}
\right]^{1/3},
\label{eq:RHII}
\end{equation}
where $\dot N_{\rm ph}$ is the ionizing photon production rate of the
central source, $\alpha_B$ is the hydrogen recombination coefficient
evaluated at $\approx 10^4$K, and $\langle n_{\rm H}^2 \rangle$ is the
volume averaged mean squared density of cold hydrogen within $R_{\rm
HII}$.  Note that in our models, the cold gas is compressed by a
factor of $\eta_{\rm cold}$, which enhances the local recombination
rate by $\eta_{\rm cold}^2$. However, the cold gas occupies only a
fraction $f/\eta_{\rm cold}$ of the volume, and hence the compression
increases the total recombination rate by an overall factor of
$f\eta_{\rm cold}$.

Assuming a fixed cold mass fraction $f=0.5$ across the halo, and
assuming further that the central BH shines at the Eddington
luminosity (for typical quasar spectra, this corresponds to an
ionizing photon production rate of $\dot N_{\rm ph}\approx
6\times10^{47}$ photons s$^{-1}$ per ${\rm M_\odot}$ of BH mass; see
Cen \& Haiman 2000), we then find the required size of the BH so
that the Str\"omgren sphere extends all the way out to the virial
radius:
\begin{equation}
M_{\rm bh}\approx 6\times 10^8~{\rm M_\odot}
\left(\frac{M_{\rm halo}}{10^{12}{\rm M_\odot}}\right)^{5/3}
\left(\frac{1+z}{6}\right)^{4}.
\label{eq:RsRhalo}
\end{equation}
Here we have utilized a relation between the halo mass, radius, and
virial temperature from NFW.  Equation~(\ref{eq:RsRhalo}) can be
understood by recalling the scalings $R_{\rm HII}^3 \propto M_{\rm bh}
\langle n_{\rm H}^2 \rangle^{-1} \propto M_{\rm bh} T_{\rm vir}^{-1}
(1+z)^{-6}$; $T_{\rm vir}\propto M_{\rm halo}^{2/3} (1+z)$; and
$R_{\rm vir}\propto M_{\rm halo}^{1/3} (1+z)^{-1}$.

Equation~(\ref{eq:RsRhalo}) reveals that in most halos of interest
($3\lsim z\lsim 8$; $10^{10}~{\rm M_\odot}\lsim M_{\rm halo}\lsim
10^{12}~{\rm M_\odot}$), converting $\lsim2\%$ of the gas mass into a
central BH is sufficient to keep most of the cold gas photoionized.
This conclusion is conservative, since we assumed a constant cold
fraction $f=0.5$, which maximizes the internal clumping and the total
recombination rate (cf.  Eqn.\ref{eq:frac}).  Using the profiles of
$f(r)$ obtained in our models, we have verified that for all halos
considered in this work, the cold gas can be fully ionized by still
smaller BHs, in all cases with $M_{\rm bh}\leq 0.01\times
(\Omega_b/\Omega_m) M_{\rm halo}$.  Although the masses of BH's in
protogalaxies at $z\gsim 3$ are unknown, ratios as large as $1\%$ of
the gas mass would be consistent with the sizes of supermassive BH's
found in nearby galaxies (Magorrian et al. 1998; Gebhardt et al. 2000;
Ferrarese \& Merritt 2000), especially since this ratio can evolve
(and decrease) as a function of redshift (Menou, Haiman \& Narayanan
2000).

\subsection{Surface Brightness and Angular Size of Ly$\alpha$ Fuzz}

Assuming that all of the cold gas in the model halo is photoionized,
we obtain the total Ly$\alpha$ line luminosity as
\begin{equation}
L_\alpha = 0.68\times E_\alpha \int_0^{R_{\rm vir}} 
4\pi r^2 dr
\left(\frac{f}{\eta_{\rm cold}}\right)
n_{\rm H}^2 \alpha_B,
\label{eq:lya}
\end{equation}
where $E_\alpha=10.2$eV is the energy of a Ly$\alpha$ photon, and the
integral represents the total recombination rate of photoionized gas
within the halo.  We have explicitly included the volume--filling
factor $(f/\eta_{\rm cold})$ of the cold gas, whose hydrogen number
density is $n_{\rm H}=0.76\rho_{\rm cold}/m_p$, and the fraction 0.68
of case B recombinations that yield a Ly$\alpha$ photon (Osterbrock
1989).  We also note that additional recombination lines might be
observable, including those of heavy elements, provided the halo gas
is sufficiently pre--enriched (see De Breuck et al. 2000 for a review
of relevant observations).

\myputfigure{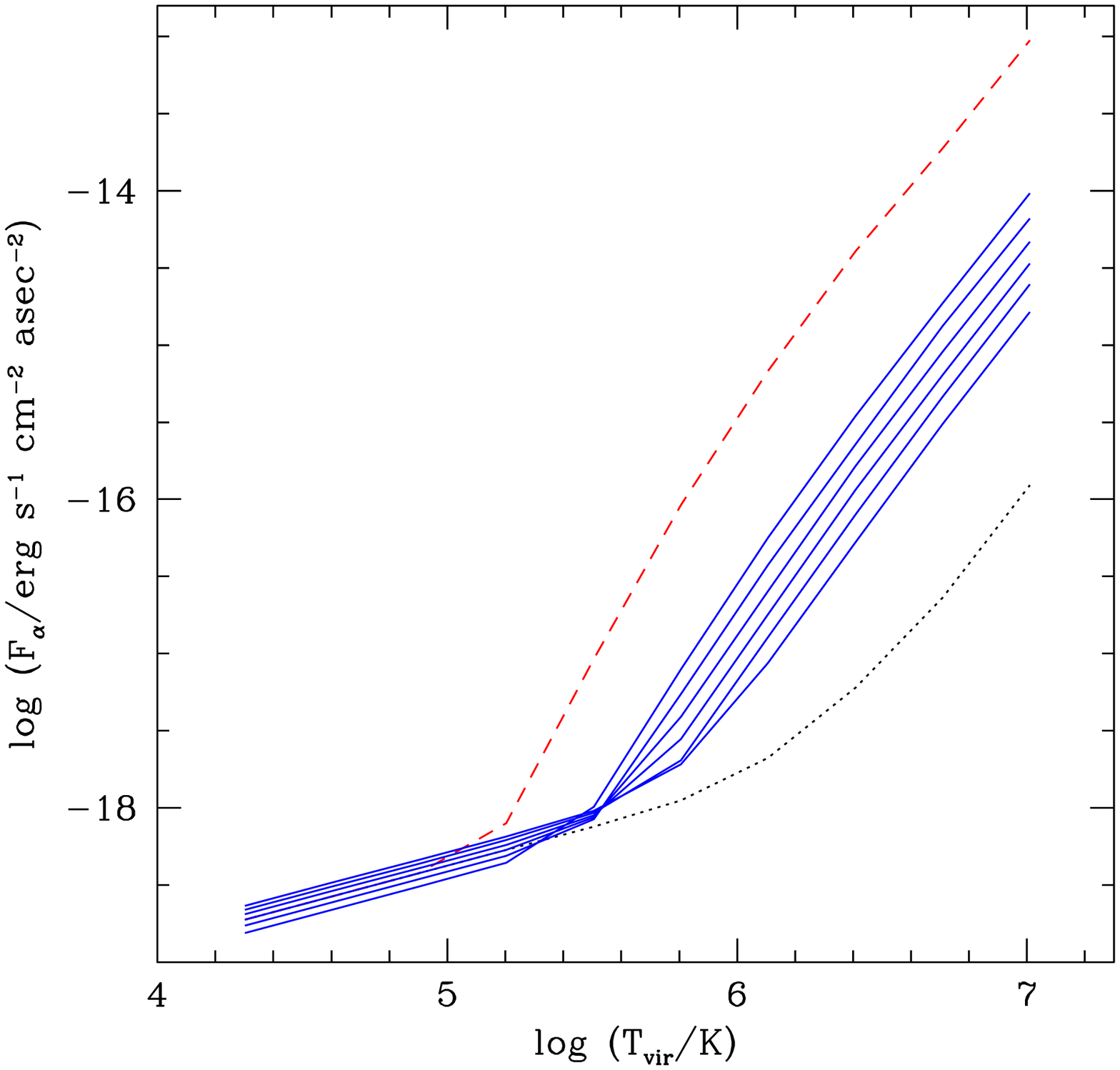}{3.2}{0.45}{-10}{-10} \figcaption{\label{fig:lyaz}
The characteristic surface brightness of Ly$\alpha$ fuzz around halos
as a function of virial temperature.  The solid curves correspond to
redshifts $z=3,4,5,6,7,8$ (top to bottom near $10^7$K).  The dashed
and dotted curves describe models (both at $z=5$) in which the cooling
time is increased by a factor of 10, or the age of the system is
increased by a factor of 5, respectively.}
\vspace{\baselineskip}

For the halos shown in Figure~\ref{fig:fcold}, the values of
$L_\alpha$ are $7\times 10^{41}~{\rm erg~s^{-1}}$, $3\times
10^{42}~{\rm erg~s^{-1}}$, $4\times 10^{43}~{\rm erg~s^{-1}}$, and
$3\times 10^{46}~{\rm erg~s^{-1}}$ (at z=3), and $10^{42}~{\rm
erg~s^{-1}}$, $5\times 10^{42}~{\rm erg~s^{-1}}$, $5\times
10^{43}~{\rm erg~s^{-1}}$, and $4\times 10^{46}~{\rm erg~s^{-1}}$ (at
z=5).  The spatial extent of the Ly$\alpha$--emitting gas is a
fraction of the virial radius $R_{\rm vir}\sim 10-100$ kpc.  As
Figure~\ref{fig:fcold} shows, this fraction varies inversely with the
virial temperature of the halo.  Here we define $R_{1/2}$ as the
radius at which the cold fraction is $f=0.5$, and assume that the
Ly$\alpha$ radiation is emitted from within a characteristic radius,
which we take to be $R_\alpha\equiv$ min$(R_{1/2},R_{\rm vir})$.  The
characteristic angular size $\theta_\alpha$ of the
Ly$\alpha$--emitting fuzz should then be
$\sim\theta_\alpha=R_\alpha/d_A$, where $d_A$ is the angular diameter
distance. We find that for the halos shown in Figure~\ref{fig:fcold},
$\theta_\alpha\approx 2-3$ arcseconds.  The angular size varies
relatively little with redshift and virial temperature, and implies
that the typical Ly$\alpha$--emitting fuzz should appear extended and
can be resolved with optical instruments.

The characteristic Ly$\alpha$ surface brightness within
$\theta_\alpha$, $F_\alpha=L_\alpha/4\pi d_L^2 /\pi
\theta_\alpha^2$ (where $d_L$ is the luminosity distance), is shown in
Figure~\ref{fig:lyaz} as a function of $T_{\rm vir}$.  The solid
curves correspond to redshifts $z=3,4,5,6,7$, and $8$ (top to bottom
near $10^7$K).  This figure reveals several interesting features.
First, the surface brightness is generally high, and all but the
smallest halos should be potentially detectable with current
instruments (see discussion below, and in Haiman, Spaans \& Quataert
2000).  Second, nearly all of the halo gas remains cold upto a virial
temperature of $\sim 3\times 10^5$K (cf. Fig.~\ref{fig:fcold}).  In
this case, the Ly$\alpha$ fuzz extends out to the virial radius, and
the surface brightness has the simple scaling $F_\alpha\propto \rho^2
V / R_{\rm vir}^2 (1+z)^4
\propto T_{\rm vir}^{1/2} (1+z)^{1/2}$ (here $V\propto R_{\rm vir}^3$
is the volume of the halo); i.e. the surface brightness weakly
increases with both virial temperature and redshift. The latter
result arises primarily from the strong dependence of the recombination
rate on redshift, $\rho^2\propto(1+z)^6$ [vs. the $(1+z)^4$ surface
brightness dimming].  Third, for higher virial temperatures ($\gsim
3\times 10^5$K), the cold fraction $f$ falls somewhat below unity,
allowing the cold phase to be compressed, and enhancing the
recombination rate and the surface brightness.  In this range of
virial temperatures, the surface brightness scales approximately as
$F_\alpha\propto T_{\rm vir}^{5/2} (1+z)^{-2}$ [note that
$\Lambda(T)\propto T^{1/2}$].  Both the stronger dependence on $T_{\rm
vir}$ and the inverse scaling on redshift result from the dependence
of the cold fraction on $T_{\rm vir}$.  Note that for still higher
virial temperatures, the cold fraction would decrease to negligibly
small values, and the surface brightness would drop sharply, but this
happens only for exceedingly large halos.

In order to illustrate the robustness of the above conclusions, we have
computed the surface brightness in two variants of our model.  First, to
facilitate comparison with earlier work, we adopt $t_{\rm cool}=t_{\rm Hub}$
rather than $t_{\rm cool}=0.2 t_{\rm Hub}$, as the condition used to compute
the cold fraction $f$.  The effect of this change on the surface brightness at
$z=5$ is shown by the dotted curve in Figure~\ref{fig:lyaz}.  As expected, the
cold fraction remains unity for halos upto a higher virial temperature, which
reduces the surface brightness by an order of magnitude at $T_{\rm vir}\gsim
10^6$K.  Second, we note that our simplified model would conflict with
observations if we applied it to local galaxy clusters (see, e.g. Fabian 1994
for a review).  Our model would predict a non--negligible amount of cold gas at
the center of at least the lowest mass clusters (e.g. $M\sim 10^{14}~{\rm
M_\odot}$ at redshift $z\sim 0$), where observations show little evidence for
cold gas.  This could mean that, in clusters, the cool gas converts quickly and
efficiently into (low--mass) stars. Alternatively, there could be heat input
(e.g.  from higher--mass star formation, SNe, etc.)  that effectively slows
down the cooling.  To mimic this latter scenario, we have increased the cooling
time by a factor of 10, equivalent to requiring $t_{\rm cool}=0.02 t_{\rm
Hub}$.  This condition ensures that $f\sim 0$ is predicted for all $z\sim 0$
galaxy clusters.  The effect of this change (at $z=5$) is shown by the dashed
curve in Figure~\ref{fig:lyaz}.  The cold fraction is decreased, which enhances
the surface brightness by upto an order of magnitude for halo with $T_{\rm
vir}\gsim 10^5$K.

\section{Observational Prospects: Constraints on Galaxy and Quasar Formation}
\label{sec:constraints}

The main results of the previous two sections are (i) in galaxy--sized
halos ($T_{\rm vir}\gsim 10^6$K) at redshifts $z\gsim 3$, a
significant fraction of the gas should be cold; and (2) if this gas is
illuminated by ionizing radiation from a central quasar, it should be
kept photoionized, and result in a detectable Ly$\alpha$ fuzz of
characteristic size of $\gsim 2$ arcseconds, and surface brightness
$\gsim 10^{-17}~{\rm erg~s^{-1}~cm^{-2}~asec^{-2}}$.

A handful of existing observations in various other contexts have
already reached these relevant levels for the Ly$\alpha$ surface
brightness.  Examples are the imaging of a proto--cluster region of
Lyman Break Galaxies at redshift $z\approx 3$ in a narrow--band
Ly$\alpha$ filter, reaching a sensitivity of $\sim 10^{-17}~{\rm
erg~s^{-1}~cm^{-2}~asec^{-2}}$ in a $\sim$16 hour observation with the
Palomar 200--inch telescope (Steidel et al. 2000).  Similar
sensitivities have been reached in $\sim$5 hours with the Keck
telescope in blank field searches for high--redshift Ly$\alpha$
galaxies (Cowie \& Hu 1998), and should also be achievable by
Subaru.\footnote{See http://www.subarutelescope.org} Extended
Ly$\alpha$ emission has also been detected around high--redshift radio
galaxies (e.g. De Breuck et al. 2000) and radio--loud quasars
(possibly related to outflows; e.g. Heckman et al. 1991a,b; Bremer et
al. 1992).  Nevertheless, observations have not yet targeted bright
quasars to search for extended Ly$\alpha$ emission at similar depths.
Less sensitive observations (reaching a few $\times 10^{-16}~{\rm
erg~s^{-1}~cm^{-2}~asec^{-2}}$) did target bright quasars.  Bremer et
al. (1992) finds extended emission around two out of three $z>3$
quasars, but a search of another twelve quasars at similar redshifts
revealed emission only around one source (Hu \& Cowie 1987).  Yet
another search around a $z=4.7$ quasar (Hu et al. 1996) has uncovered
discrete Ly$\alpha$ emitting companions, rather than an extended
continuous ``fuzz''.

In summary, the few existing observations, typically utilizing $\sim1$
hour integrations on 4m telescopes, probe surface brightnesses above
$\gsim 10^{-16}~{\rm erg~s^{-1}~cm^{-2}~asec^{-2}}$, and reveal that
luminous quasars are not generally enveloped by Ly$\alpha$ fuzz at
this sensitivity.  This puts a mild constraint on the models, although
emission at this level is expected only in the largest halos (see
Fig. \ref{fig:lyaz}).  However, observations that probe bright quasars
at about an order of magnitude deeper, would either detect Ly$\alpha$
fuzz around most sources, or else lead to strong constraints on the
type of models discussed here.  Provided that the Ly$\alpha$ fuzz {\it
is} detected, its surface brightness, shape and extent, together with
the line--profiles, would constitute an invaluable direct probe of
galaxy formation.

It is interesting to consider constraints that would arise on galaxy
and quasar formation, should the current trend of {\it not} detecting
fuzz persist at fainter fluxes.  One possibility is that quasars turn
on only during the later stages of galaxy formation, i.e. at a time
when most of the cold gas has already settled to a thin disk, and/or
turned into stars. A lack of Ly$\alpha$ fuzz would then be naturally
explained by the absence of significant amounts of spatially extended
cold gas during the luminous quasar phase.  To avoid detectability at
the surface brightness threshold of $\sim 10^{-17}~{\rm
erg~s^{-1}~cm^{-2}~asec^{-2}}$, Figure~\ref{fig:lyaz} suggests that in
halos with $T_{\rm vir}\gsim 10^6$K, the flux has to be reduced by a
factor $\gsim 10$ relative to the predictions of our simple models.
This, implies, in turn, that $\gsim 90\%$ of the cold gas must already
have settled to a disk (or disappeared).  It has been argued that the
presence of a disk is indeed a pre--requisite for the central BH to
grow (Sellwood \& Moore 1999); furthermore, cold gas in a thin disk is
locally Toomre--unstable, and might rapidly turn into stars (e.g. Mo
et al. 1998).  Independent ``evidence'' favoring this scenario is that
the heavy element abundances in the broad line region of quasars
always appear to be high (Hamann 1999), even at high redshifts.  This
requires one or two generations of massive stars to precede the
activation of the central quasar.

There are other, less attractive scenarios without Ly$\alpha$
emission. If the cold gas is quickly turned into stars, and/or
collapse to the central regions, or to a thin disk, this could
eliminate Ly$\alpha$ reprocessing, at least until cold gas is
replenished by further accretion or merger with another halo
(i.e. resulting in a short Ly$\alpha$ duty--cycle).  However, this
would also imply that a fraction $\sim t_{\rm dyn}/t_{\rm Q}\gsim
10\%$ of all quasars should still show Ly$\alpha$ fuzz, unless the
cold gas disappears on an exceedingly short timescale $(\ll t_{\rm
dyn})$ (see Haiman \& Hui for constraints on the lifetime $t_Q$ of the
luminous quasar phase).  Alternatively, one might envision that the
cold gas resides in clumps with a small covering factor, allowing most
of the ionizing photons to leak out along line of sights traversing
only hot (collisionally ionized) medium.  However, this explanation
requires a minimum cold clump size that exceeds the Jeans mass in the
cold phase (Rees 1988). Hence, the postulated large clumps are
unstable and would fragment to smaller pieces, increasing the covering
factor to approximately unity. Yet another scenario with no Ly$\alpha$
fuzz is if the quasar (or its associated wind) has blown out most of
the gas from the halo.  However, in this case, one would still expect
to see a fluorescent Ly$\alpha$ outflow around a fraction of quasars
``caught in the act'' of removing the gas.

Dust absorption might strongly suppress the Ly$\alpha$ flux escaping from a
medium, even if the medium is optically thin in the Lyman continuum, and has
been thought to cause the lack of detections of proto--galaxies in early
Ly$\alpha$ surveys. A strong quenching of Ly$\alpha$ by dust, however, does not
necessarily occur (see, e.g. Neufeld 1991, or Pritchet 1994 for a general
discussion), especially if most of the Ly$\alpha$ photons originate in an
ionized layer with relatively low dust opacity.  Furthermore, Ly$\alpha$
emitting galaxies have been found at high--redshift (e.g. Hu et al. 1996;
1998), as expected in models with lower galactic dust abundance, and
inhomogeneous dust distribution (Haiman \& Spaans 1999).  Based on these
observations, it would appear unlikely that dust can suppress the Ly$\alpha$
fuzz from around all high--redshift quasars. Indeed, the dust abundance in the
early, spatially extended, collapsing phase of the high--redshift halos is
likely to be significantly lower than in star--forming galaxies.

Even if dust does not suppress the Ly$\alpha$ emission itself,
however, an important question is whether the nucleus can be obscured
by dust and rendered undetectable at optical wavelengths.  If this is
typical, then Ly$\alpha$ fuzz should be expected around sub-mm
sources, rather than around optical quasars.  The extended Ly$\alpha$
emitting blob of Steidel et al. (2000) has been found to be an
exceptionally bright sub-mm source (Chapman et al. 2000), with no
visible continuum source.  The Ly$\alpha$ line luminosity for this
object is $\sim 2\times10^{10}{\rm L_\odot}$, while the bolometric
luminosity inferred from the sub--mm detections is $>10^{13}{\rm
L_\odot}$.  This implies that less than $1\%$ of the UV produced by
the source is available to power the extended Ly$\alpha$ emission, and
that most of this UV is used up for this purpose (to explain the lack
of any continuum source).  It would be surprising if extended
Ly$\alpha$ fuzz was typically produced in a similar way, since it
requires a fine--tuning of the unobscured fraction to match the amount
of surrounding cold gas.  Nevertheless, it would be invaluable to
target bright sub--mm sources (i.e. those with redshift estimates) in
deep Ly$\alpha$ searches, to clarify what fraction of them do produce
Ly$\alpha$ emission.  The primary driver (stellar vs. quasar UV light)
of both the sub-mm and Ly$\alpha$ emissions remain unclear in the
Chapman et al. (2000) source (as well as in other bright sub-mm
sources). Nevertheless, the lack of strong sub-mm emission in a second
Ly$\alpha$ blob in the same dataset suggests that the line emission is
not necessarily powered by a dust--obscured source.

Our models imply a second interesting, although somewhat less
stringent constraint on the amount of extended cold gas around
quasars.  As discussed in \S~3.1 above, we find that it is sufficient
to convert $\sim 1\%$ of the total gas mass into a central BH in order
to keep most of the cold gas ionized (cf. Eqn.~\ref{eq:RsRhalo}).  We
emphasize that this must indeed happen if any ionizing radiation is to
escape from the halo.  Observations typically indicate that a large
fraction of the ionizing radiation from quasars does escape, even for
the quasar with the highest known redshift at $z=5.8$ (Fan et
al. 2000b).  The abundance of this object implies a halo mass $M_{\rm
halo}\sim 10^{13}~{\rm M_\odot}$, while its luminosity, under the
assumption that it equals the Eddington limit, implies a BH mass of
$M_{\rm bh}\sim 4\times 10^9~{\rm M_\odot}$ (Haiman \& Loeb 2000).
For this halo, under the assumption of a constant cold fraction
$f=0.5$, equation~(\ref{eq:RsRhalo}) would imply that a BH as large as
$M_{\rm bh}\sim 4\times 10^{10}~{\rm M_\odot}$ is needed to ionize the
cold gas and allow the ionizing radiation to escape.  Based on the
profile $f(r)$ derived in our models, the requisite BH mass in
smaller, $M_{\rm bh}\sim 1.3\times 10^{10}~{\rm
M_\odot}$. Nevertheless, this mass is close to $1\%$ of the total gas
mass, and is a factor of $\sim$ two higher than the BH mass inferred
directly from the luminosity.  We conclude that a large escape
fraction of the ionizing continuum from bright, high--redshift quasars
requires either (1) massive central BH's whose masses are a
significant fraction ($\sim 1\%$) of the gas mass, or (2) that some of
the cold gas has settled to a disk (or disappeared in a blow--out).

Finally, we note that any significant continued gas infall at large
radius, or a major merger spreading cold gas over an extended region,
could still reprocess much of the quasar's radiation to Ly$\alpha$;
although the surface brightness at large radii might drop below
detectable levels.

\section{Conclusions}
\label{sec:conclusions}

In this paper, we have studied what would happen to a protogalaxy if a
bright quasar were to turn on during the early stages of its assembly.
Using a simple spherical model for the distribution of a two--phase
gas, we find that such a system should have a substantial amount of
cold, photoionized gas with a spatially extended distribution.  We
expect this conclusion to be generic, due to efficient radiative
cooling at high redshifts.  The cold, photoionized gas is detectable
as an extended Ly$\alpha$ fuzz enshrouding the quasar; with a
characteristic angular size of a few arcseconds, and a surface
brightness of $\sim 10^{-18}-10^{-16}~{\rm
erg~s^{-1}~cm^{-2}~asec^{-2}}$.  Existing observations have not yet
targeted luminous quasars at these sensitivities, although at $\sim
10^{-16}~{\rm erg~s^{-1}~cm^{-2}~asec^{-2}}$, most quasars do not
appear to be enveloped by extended emission.  Future, deep Ly$\alpha$
imaging of few arcsecond regions around luminous quasars (as well as
bright sub--mm sources) should be possible with current instruments.
While a detection of Ly$\alpha$ fuzz would provide a direct probe of
galaxy formation; non--detections at the level of $10^{-17}~{\rm
erg~s^{-1}~cm^{-2}~asec^{-2}}$ would already imply strong constraints.

\acknowledgements

We thank Eliot Quataert for useful discussions, and an anonymous
referee for comments that improved this paper.  This work began when
MJR was visiting Princeton University as Scribner Lecturer; he
acknowledges support from this source, and from the Royal Society. ZH
was supported by NASA through the Hubble Fellowship grant
HF-01119.01-99A, awarded by the Space Telescope Science Institute,
which is operated by the Association of Universities for Research in
Astronomy, Inc., for NASA under contract NAS 5-26555.

\end{document}